# A Formal Physical Framework for the Origin of Life: Dissipation-Driven Selection of Evolving Replicators

Author: Shlomo Segal

## Abstract

The emergence of life from inanimate matter presents a thermodynamic challenge: the Second Law of Thermodynamics dictates a global trend towards disorder, yet life constitutes localized pockets of profound organization. This paper presents a formal physical framework for abiogenesis grounded in the statistical physics of non-equilibrium systems. We transition from the established connection between dissipation and process probability (e.g., Crooks Fluctuation Theorem) to a large-deviation framework for the likelihood of system histories. This formalism reveals a probabilistic bias towards histories with greater integrated dissipation. We then demonstrate how this bias leads to the selection of heredity. The core of our argument is a rigorous mathematical proposition showing that while simple autocatalysis leads to an exponential increase in dissipation, template-directed replication, via its capacity for mutation and adaptation (a process from which we derive an effective adaptation rate, $\alpha$), unlocks a super-exponential growth pathway. This translates to a doubly-exponential amplification in the relative probability of its emergence over time, constituting an asymptotically dominant physical bias for its selection. This framework delineates a hierarchical transition from simple dissipative structures to information-bearing replicators, whose stability is contingent upon exceeding critical thresholds of fidelity, kinetic efficiency, and resource supply. We conclude by proposing a refined, quantitative, and falsifiable experiment, defining a precise mathematical signature for identifying the onset of evolutionary processes in synthetic chemical systems.

## 1. Introduction: Thermodynamics and the Emergence of Life

The existence of life, a localized state of extraordinarily low entropy, appears to contend with the Second Law of Thermodynamics. Erwin Schrodinger's (1944) seminal insight that living systems maintain order by feeding on "negative entropy"—that is, by exporting entropy to their environment—resolved the apparent paradox. However, this does not explain the origin of such complex, information-processing systems. Our work addresses this gap by proposing a physical selection principle that favors the emergence of heredity.
We do not assume a simple maximization of entropy production. Instead, we leverage the principles of modern statistical physics for non-equilibrium systems to argue for a probabilistic bias favoring system histories that integrate to a greater total dissipation over time. Our central goal is to demonstrate mathematically the conditions under which this bias leads not just to generic order, but specifically to the emergence of information-bearing, evolvable replicators.

## 2. A Formalism for Dissipation-Driven Selection

The foundation of our argument is the statistical physics of non-equilibrium processes. The Crooks Fluctuation Theorem relates the probability of a microscopic trajectory to its time-reverse via the total entropy produced. This theorem rigorously connects dissipation with the statistical likelihood of paths. In the limit of long durations ($\tau$), the probability of a particular macrostate trajectory can be expressed as an exponential function of a rate functional. We make the central ansatz of this work, motivated by the underlying connection between dissipation and irreversibility, that the likelihood of two distinct forward histories, $x1$ and $x2$, over a duration $\tau$ can be compared via their total dissipative outputs:

$$\frac{P(x1)}{P(x2)} \text{ is approximately equal to } exp(\frac{(\sigma 1 - \sigma 2)}{2})$$

where $\sigma$ is the dimensionless total entropy produced along that history, defined as the integral of the heat dissipation rate divided by temperature and the Boltzmann constant. This framework allows for a direct probabilistic comparison of histories based on their total dissipative output.

**2.1. Mathematical Model: The Probabilistic Selection of Replication**

Consider a chemical system fed by high-energy fuel molecules F.
In a physically realistic driven chemical system, the replication kinetics must depend explicitly on the concentration of the fuel molecules F, which provide the free energy required to sustain the non-equilibrium state. The effective replication rate is therefore more accurately written as a function of fuel availability, k(F). For simplicity, and to maintain analytical transparency, we assume that the system operates in a regime where the fuel supply is maintained by a continuous external flux, allowing F to remain approximately constant over the timescale considered.
We compare two idealized histories starting from a small number of initial molecules (n0): one dominated by simple autocatalysis (History A) and one by adaptive, template-based replication (History R).

**Step 1: Population Growth Dynamics**

History A (Simple Autocatalysis): A molecule A catalyzes its own formation from fuel F. The kinetics follow the equation $\frac{dn}{dt} = k_A \cdot n$ leading to simple exponential growth: $n(t) = n_0 \cdot exp(k_A \cdot t)$.

History R (Adaptive Replicator): A molecule R replicates via a template mechanism. The capacity for adaptation means the effective mean kinetic parameter of the population, $k(eff)$, increases over time as more efficient sequences are discovered. This is approximated as a linear improvement: $k(eff) = k_0 + \alpha \cdot t$ This linear increase should be understood as a first-order approximation valid during the early adaptive phase.
In realistic systems the improvement in replication kinetics is expected to saturate due to biochemical and physical constraints, approaching a maximal rate k_max. The linear form is therefore used only as a minimal analytic model capturing the initial regime of adaptive improvement.

where $\alpha$ is the rate of adaptation. The kinetics follow $\frac{dn}{dt} = n(k_0 + \alpha \cdot t)$ leading to super-exponential growth: $n(t) = n_0 \cdot exp(k0 \cdot t + (\frac{\alpha}{2}) \cdot t^2)$.
The $t^2$ term in the exponent drives this accelerated growth.
Integrating the kinetic equation $\frac{dn}{dt} = n(k_0 + \alpha \cdot t)$ yields the population trajectory $n(t) = n_0 \, exp(k_0 t + (\frac{\alpha}{2}) t^2)$.

**Step 2: Integrated Dissipation**

The total entropy produced, $\sigma$, is the time integral of the entropy production rate, which is proportional to the reaction rate. For History A, the total dissipation grows exponentially. For History R, the total dissipation grows faster than exponentially due to the accelerating replication dynamics. proportional to $exp(k_0 \cdot \tau + (\frac{\alpha}{2}) \cdot \tau^2)$.

**Step 3: The Probabilistic Selection**

We define the selection ratio Gamma as the ratio of probabilities $\frac{P(R)}{P(A)}$. Substituting the expressions for dissipation, we find that Gamma grows doubly-exponentially with time: $exp(\,exp(\,\tau^2\,)\,)$.
 The dominance of the replicative history is assured for sufficiently large $\tau$, as the super-exponential term causes History R to overpower History A. This constitutes an asymptotically dominant probabilistic bias toward the system history that discovers heredity.

## 2.2. Proposition 1

In a driven chemical system containing precursors for both a static autocatalyst and an adaptable replicator, the ratio of the probabilistic weight of the replicative history to the autocatalytic history grows doubly-exponentially for large $\tau$. Asymptotically, the log of the log of Gamma is proportional to $(\frac{\alpha}{2}) \cdot \tau^2$. Furthermore, the replicative pathway's dissipation rate will dominate any purely exponential pathway as time approaches infinity.

### 2.2.1. Critical Takeover Time

The super-exponential pathway begins to outpace the exponential one only after a characteristic time, $\tau(crit)$. This occurs when the cumulative effect of adaptation ($\alpha$) overcomes any initial kinetic disadvantage of the replicator system. If the initial rate of the replicator is lower than the autocatalyst, a specific duration is required for the "intelligence" of the system to compensate for its initial slowness.

## 2.3. The Microscopic Origin of the Adaptation Rate $\alpha$

The parameter $\alpha$ emerges from a minimal model of mutation and selection. Fisher's fundamental theorem of natural selection states that the rate of increase in mean fitness is equal to the genetic variance in fitness. For a system with a constant mutation rate exploring a landscape with a selective gradient, the variance can be sustained for a period, leading to a quasi-linear increase in the mean replication rate. Here, $\alpha$ is an emergent parameter encapsulating the mutation rate, the length of the replicator, and the local topology of the fitness landscape.

## 3. Necessary Threshold Conditions for Replicative Takeover

The selective advantage outlined in our proposition is conditional upon clearing critical physical thresholds:

1. Fidelity Threshold (Eigen's Error Threshold): Replication must be accurate enough to pass information. If the mutation rate is too high, $\alpha$ becomes zero or negative, erasing heritable advantage and collapsing the system back to a non-evolving state.
2. Kinetic Threshold: The initial replicator must have a replication rate that outpaces its natural decay or degradation rate.
3. Resource Threshold: The framework is contingent on a sustained high chemical potential of the fuel source. Resource depletion halts growth and dissolves the non-equilibrium state.
4. Parasitism: The replicative system must remain kinetically superior to "parasitic" sequences—shorter strands that replicate faster by hijacking machinery but providing no functional advantage.

## 4. The Hierarchical Emergence of Life

This framework suggests a hierarchical sequence of origins: Level 1: Basic Dissipative Structures (e.g., convection cells), which achieve spatial order but lack memory.
Level 2: Autocatalytic Chemical Networks, which are capable of exponential dissipation but lack open-ended evolution. Level 3: Information-Bearing Replicators, where thermodynamics begins to select for systems capable of Darwinian evolution.

## 5. Evolutionary Selection as a Thermodynamic Consequence

It is important to emphasize that this framework does not imply any violation of the Second Law. Living systems maintain low internal entropy only by exporting larger amounts of entropy to their surroundings. The emergence of replicators therefore increases the total entropy production of the combined system and environment. The thermodynamic bias described here thus reflects the tendency of driven systems to evolve structures that dissipate free energy more efficiently. Our framework allows for complex strategies.
A system might temporarily follow a path of lower dissipation if that path leads to the discovery of an innovation that unlocks a far greater total dissipation over the entire history. The selection is for the most dissipative history, not necessarily the most dissipative instant. This provides a thermodynamic underpinning for "Dynamic Kinetic Stability". In a driven environment, the most persistent replicators are those that are kinetically most proficient.

## 6. Experimental Design: Testing the Thermodynamic Signature of Life

We propose that any chemical environment leading to the emergence of evolving replicators will be identifiable by a specific mathematical signature.

Experimental Setup: A continuous-flow reactor monitored by isothermal calorimetry, fueled by activated monomers (e.g. RNA monomers) and seeded with a diverse pool of random sequences. Predicted Signature: On a semi-log plot of the log of power dissipation versus time, the takeover by an evolving system will appear as a transition from a linear regime to a distinctly convex curve.

The key experimental signature is: $\frac{d^2}{dt^2}(log(P_{diss})) > 0$.

## 7. Conclusion

We have presented a formal physical argument connecting the statistical physics of non-equilibrium systems to the origin of heredity. The acquisition of heritable information is a chemical strategy that unlocks a super-exponentially growing dissipative capacity. Darwinian evolution is thus cast not as a novel force appearing from nothing, but as the macroscopic expression of a trajectory overwhelmingly favored by the probabilistic laws governing driven matter once critical thresholds are surpassed.

## 8. Model Assumptions and Limitations

This model assumes a non-depleting resource supply and neglects the potential impact of spatial heterogeneity or compartmentalization. Future research should integrate coupled differential equations for different species to model parasite dynamics more accurately.